\documentclass[9pt]{article}
\usepackage{spconf,amsmath,graphicx,hyperref}
\usepackage{amsfonts}
\usepackage{xcolor}
\usepackage{cite}

\title{Tracking Dynamic Simplicial Complexes \\ via Constrained State-Space Estimation}
\name{Varun Sarathchandran, Bishwadeep Das, Elvin Isufi and Geert Leus \thanks{Emails: v.sarathchandran@tudelft.nl, b.das@tudelft.nl, \\ e.isufi-1@tudelft.nl, g.j.t.leus@tudelft.nl}}

\address{Delft University of Technology, the Netherlands }
\begin{document}
\ninept

\maketitle
\begin{abstract}

Simplicial complexes (SCs) extend graphs to represent higher-order interactions, but couple their simplex levels through the inclusion property. While static SC inference is an emerging research direction, tracking time-varying SCs remains largely unexplored. A central challenge in tracking SCs is to account for the inclusion property. To this end, we build a nonlinear state-space model tailored to SCs. For prediction, we introduce a closure-aware Markov generation model whose conditional mean preserves simplicial inclusion and whose covariance captures edge-triangle dependencies. For correction, we encode inclusion constraints as nonlinear pseudo-measurements, allowing the constraints to inform both the state and covariance updates. We investigate progressively richer treatments of these measurements through a standard extended Kalman filter, an iterated extended Kalman filter, and a Laplace approximation. Experiments then demonstrate the benefits of exploiting the proposed dynamics and constraint information.
\end{abstract}
\section{Introduction}
\label{sec:intro}
Graphs represent pairwise relationships, whereas many natural systems involve interactions among groups of entities. Simplicial complexes (SCs) model such higher-order interactions while coupling simplex levels through the inclusion property \cite{isufi2025topological}. Since these interaction structures are rarely known a priori, their inference from observed data is an important problem. Existing SC-learning methods predominantly assume static topologies \cite{isufi2025topological}.

 However, higher-order interactions may themselves evolve over time. In a social network, for example, groups of interacting individuals form simplices whose composition changes continuously. Topology inference then becomes a tracking problem. Additionally, the structures at multiple simplex levels must be recovered sequentially while satisfying SC constraints.

Recent work on tracking dynamic graphs has been centred around state-space models (SSMs). Approaches include an extended Kalman filter with sparsity-regularized state updates 
\cite{dabush2025sparsity}, an SSM that tracks beliefs over possible neighborhoods of every node \cite{tenorio2025tracking}, and particle filtering \cite{azoulay2026dynamic,cui2024inference,ancherbak2015time}. Dynamic SC inference, however, remains comparatively unexplored. Recent methods adaptively learn only triangles while filtering signals \cite{liu2026topological,marinucci2026topological}, or infer higher-order cells from known lower-order structure through reinforcement learning methods, with experiments focussing on static SCs \cite{Canbolat2025OnlineStructures}. To the best of our knowledge, no existing method jointly tracks both edges and triangles of a time-varying SC.

We address this gap through a constrained nonlinear SSM framework. To this end, building on existing dynamic SC generation models \cite{thoppe2016evolution,owada2021limit}, we first propose a closure-aware Markov model whose conditional moments yield an SC-preserving nonlinear transition  and a process covariance which captures dependencies inherent in an SC. Second, we construct a virtual-measurement model from observations by imposing signal priors commonly used in SC learning, thereby avoiding the need for known input-output signal pairs. Building on sparsity-aware filters that encode sparsity through pseudo-measurements \cite{farahmand2014tracking,carmi2009methods,hage2020sparse}, we similarly incorporate expected simplex counts into our measurement model. Third, we design constraint-aware update steps by encoding SC constraints via nonlinear pseudo-measurements \cite{simon2010kalman,husmann2025recursive}. We study progressively richer techniques to inform the covariance update with these constraints via an extended Kalman filter (EKF), iterated EKF (IEKF) and finally a Laplace approximation, each of which increasingly takes into account richer information from these constraint-imposing nonlinear pseudo-measurements.

\section{Problem Statement}

We first introduce static SCs and then extend the formulation to time-varying topologies. Given a set of $N$ nodes $\mathcal V$, an undirected $i$-simplex $\sigma^i$ is a subset of $\mathcal V$ with cardinality $i+1$. Hence, nodes, edges, and triangles are $0$-, $1$-, and $2$-simplices, respectively. An SC of dimension $D$ is a collection of simplices of dimension at most $D$ that is closed under an inclusion property, e.g., every triangle requires all three of its boundary edges. In this work, we restrict our attention to two-dimensional SCs.

Following \cite{BuciuleaLearningApproach}, we represent the SC topology through simplex selection. Let
$\bar{\mathcal E}=\{e\subseteq\mathcal V:|e|=2\}$ and
$\bar{\mathcal T}=\{t\subseteq\mathcal V:|t|=3\}$
denote the candidate edge and triangle sets, enumerating all possible edges and triangles. Thus, $|\bar{\mathcal E}|=\binom{N}{2}$ and
$|\bar{\mathcal T}|=\binom{N}{3}$. The binary vectors
$\mathbf s^1\in\{0,1\}^{|\bar{\mathcal E}|}$ and
$\mathbf s^2\in\{0,1\}^{|\bar{\mathcal T}|}$ select simplices from these candidate sets: $[\mathbf s^i]_j=1$ indicates that the $j^{\mathrm{th}}$ $i$-simplex is active. For notational convenience, each vector entry is directly indexed by its associated candidate simplex. For example, \([\mathbf s^1]_e\) denotes the entry corresponding to candidate edge \(e\). The inclusion property can then be enforced with the constraint
$[\mathbf s^2]_t\leq[\mathbf s^1]_e,
\forall t\in\bar{\mathcal T},
\forall e\in\partial t$
where $\partial t\subset\bar{\mathcal E}$ contains the three boundary edges of $t$.

Time-varying SCs follow by letting these selections change over time while keeping the node and candidate-simplex sets fixed. At time $k$, the topology is represented by
$\mathbf s_k^1\in\{0,1\}^{|\bar{\mathcal E}|}$ and
$\mathbf s_k^2\in\{0,1\}^{|\bar{\mathcal T}|}$, while the inclusion property can be guaranteed with the constraint
\begin{equation}
    [\mathbf s_k^2]_t\leq[\mathbf s_k^1]_e, \quad \forall t\in\bar{\mathcal T},\quad
\forall e\in\partial t, \quad k=1,\ldots,K.
\label{eq:time_inclusion}
\end{equation}
At each time instant, we observe nodal and candidate-edge signals collected in
$\mathbf Z_k^0\in\mathbb R^{N\times M^0}$ and
$\mathbf Z_k^1\in\mathbb R^{|\bar{\mathcal{E}}|\times M^1}$, respectively, where $M^0$ and $M^1$ denote the numbers of features at the two simplex levels. We assume that edge signals are available over the whole candidate set $\bar{\mathcal E}$. When they are not directly observed, they may instead be constructed from the corresponding nodal signals using a permutation-invariant fusion function, as in \cite{Self-DrivenPrediction}. Given a signal prior that relates these observations to the underlying SC, our goal is to sequentially infer its time-varying topology.

\noindent\textbf{Problem Statement}. Given $\{\mathbf Z_\ell^0,\mathbf Z_\ell^1\}_{\ell=1}^k$, our objective is to sequentially recover
$\mathbf s_k=[(\mathbf s_k^1)^\top,(\mathbf s_k^2)^\top]^\top
\in\{0,1\}^{|\bar{\mathcal E}|+|\bar{\mathcal T}|}$, for $k=1,\ldots,K$, subject to an assumed signal-topology relationship and the inclusion constraints in \eqref{eq:time_inclusion}.
\section{Closure-Aware SC Dynamics}

We first specify a probabilistic model for the evolution of the latent binary SC topology, which serves as the dynamical prior for our tracking methods. Let \(\beta^i\) and \(\rho^i\) be the birth and survival probabilities at simplex level \(i\), i.e., an inactive simplex is born with probability \(\beta^i\) and an active simplex survives with probability \(\rho^i\). Conditioned on
\(\mathbf s_k\), edges and intrinsic triangle proposals are sampled
independently as
\begin{equation}
\begin{aligned}
[\mathbf s_{k+1}^1]_e
&\sim\operatorname{Bernoulli}\!\left(
\beta^1+(\rho^1-\beta^1)[\mathbf s_k^1]_e\right),\\
[\mathbf u_{k+1}^2]_t
&\sim\operatorname{Bernoulli}\!\left(
\beta^2+(\rho^2-\beta^2)[\mathbf s_k^2]_t\right).
\end{aligned}
\label{eq:intrinsic_dynamics}
\end{equation}
However, a triangle
can be active only when all its boundary edges are present. After sampling all edges, triangle \(t\) is therefore gated by its new boundary:
\begin{equation}
[\mathbf s_{k+1}^2]_t
=
[\mathbf u_{k+1}^2]_t
\prod_{e\in\partial t}[\mathbf s_{k+1}^1]_e .
\label{eq:triangle_gate}
\end{equation}

Our construction builds on existing models of dynamic higher-order
structure. Dynamic clique complexes evolve edges stochastically, but
simply fill in every induced clique \cite{thoppe2016evolution}.
More generally, \cite{owada2021limit} assigns a latent on-off process to every candidate simplex and gates its realized presence by those of its boundary simplices. Because the latent triangle continues to evolve while its boundary is unavailable, it may become active as soon as boundary feasibility is restored, without undergoing a new birth. In contrast, our transition depends on the previously realized triangle state: once inactive, a triangle can reactivate only through its birth probability $\beta^2$ even if the boundary reappears. 
\section{Tracking Framework}

We use a prediction-correction framework to track $\mathbf{s}_k$ comprising a transition model that captures the SC-specific dependencies (Sec.~\ref{ssec:transition}), a measurement model that turns standard SC signal priors into virtual measurements of the observed signals (Sec.~\ref{ssec:measurement}), and update steps that respect the constraints in \eqref{eq:time_inclusion} (Sec.~\ref{sec:constrainted_update}).

\subsection{Transition Equation}
\label{ssec:transition}
To construct the filtering transition, we summarize the binary
dynamics through their first two conditional moments. For a binary previous state \(\mathbf s\), define the conditional moments
\begin{equation}
f(\mathbf s)
:=
\mathbb E[\mathbf s_k\mid\mathbf s_{k-1}=\mathbf s],
\quad
\mathbf Q(\mathbf s)
:=
\operatorname{Cov}(\mathbf s_k\mid\mathbf s_{k-1}=\mathbf s).
\end{equation}Define the edge and intrinsic triangle-transition probability vectors as
\[
\begin{aligned}
\mathbf p^1(\mathbf s)
=
\beta^1\mathbf 1_{|\bar{\mathcal{E}}|}
+(\rho^1-\beta^1)\mathbf s^1,
\mathbf p^2(\mathbf s)
=
\beta^2\mathbf 1_{|\bar{\mathcal{T}}|}
+(\rho^2-\beta^2)\mathbf s^2.
\end{aligned}
\]
The boundary-feasibility vector $\mathbf g^2(\mathbf s)$ has entries
\(
[\mathbf g^2(\mathbf s)]_t
=
\prod_{e\in\partial t}
[\mathbf p^1(\mathbf s)]_e.
\)
The conditional-mean transition is therefore
\[
f(\mathbf s)
=
\begin{bmatrix}
f^1(\mathbf s)\\
f^2(\mathbf s)
\end{bmatrix}
=
\begin{bmatrix}
\mathbf p^1(\mathbf s)\\
\mathbf p^2(\mathbf s)\odot\mathbf g^2(\mathbf s)
\end{bmatrix}.
\]
Here, $\odot$ denotes the entry-wise product. We partition the conditional covariance as
\[
\mathbf Q(\mathbf s)
=
\begin{bmatrix}
\mathbf Q^{11} & \mathbf Q^{12}\\
\mathbf Q^{21} & \mathbf Q^{22}
\end{bmatrix},
\qquad
\mathbf Q^{21}=(\mathbf Q^{12})^\top.
\]
Suppressing dependence on $\mathbf s$, and taking
$e,e'\in\bar{\mathcal E}$ and $t,t'\in\bar{\mathcal T}$, conditional
independence of the edge transitions gives,
\[
[\mathbf Q^{11}]_{e,e'}
=
\begin{cases}
[\mathbf p^1]_e\bigl(1-[\mathbf p^1]_e\bigr), & e=e',\\
0, & e\neq e'.
\end{cases}
\]
The edge-triangle covariance is non-zero only for edge-triangle incident pairs,
\[
[\mathbf Q^{12}]_{e,t}
=
\begin{cases}
[\mathbf p^2]_t[\mathbf g^2]_t
\bigl(1-[\mathbf p^1]_e\bigr),
& e\in\partial t,\\
0, & e\notin\partial t.
\end{cases}
\]
Finally, the triangle-triangle covariance can be written as
\[
[\mathbf Q^{22}]_{t,t'}
=
\begin{cases}
[\mathbf p^2]_t[\mathbf g^2]_t
\bigl(1-[\mathbf p^2]_t[\mathbf g^2]_t\bigr),
& t=t',\\[1mm]
[\mathbf p^2]_t[\mathbf p^2]_{t'}
\left(
\displaystyle\prod_{e\in\partial t\cup\partial t'}
[\mathbf p^1]_e
-
[\mathbf g^2]_t[\mathbf g^2]_{t'}
\right),
& t\neq t'.
\end{cases}
\]
Note that the off-diagonal entries of $\mathbf{Q}^{22}$ are zero when $t$ and $t'$ do not share incident edges since, in this case, $\displaystyle\prod_{e\in\partial t\cup\partial t'}
[\mathbf p^1]_e = [\mathbf g^2]_t[\mathbf g^2]_{t'}$. Thus, the structure of \(\mathbf Q(\mathbf s)\) captures both cross-level and shared-boundary dependencies induced by inclusion.

We wish to propagate these conditional moments via a transition model. However, instead of tracking the binary vector $\mathbf{s}_k$, we track its continuous extension $\mathbf{x}_k =  [(\mathbf x_k^1)^\top,(\mathbf x_k^2)^\top]^\top \in [0,1]^{|\bar{\mathcal E}|+|\bar{\mathcal T}|}$, whose entries represent soft edge and triangle activations. For $\mathbf x_k$ to represent a valid relaxed SC, we require $ \mathbf x_k\in\mathcal C$, where 
\[\mathcal C
=
\left\{
\mathbf x\in[0,1]^{|\bar{\mathcal E}|+|\bar{\mathcal T}|}:
[\mathbf x^2]_t\leq[\mathbf x^1]_e,\;
\forall t\in\bar{\mathcal T},\ e\in\partial t
\right\}.
\] 
Extending $f(\mathbf{s})$ and $\mathbf{Q}(\mathbf{s})$ to $\mathbf{x}$ yields the transition 
\begin{equation}
    \mathbf x_k
=
f(\mathbf x_{k-1})+\mathbf w_k \qquad
\mathbf w_k\sim\mathcal N(\mathbf 0, \mathbf{Q}(\mathbf{x}_{k-1})).
\end{equation}
This construction involves two modelling approximations. First, we approximate the Bernoulli transition by a Gaussian having the same conditional mean and covariance. Second, we evaluate the conditional moments, defined on the binary state, at continuous values. Thus, $f(\mathbf{x})$ and $\mathbf{Q}(\mathbf{x})$ are no longer the exact conditional moments. Instead, they parameterize a continuous, heteroscedastic Gaussian surrogate transition: $\mathbf{x}_k \mid \mathbf{x}_{k-1}=\mathbf{x} \sim \mathcal{N}(f(\mathbf{x}),\mathbf{Q}(\mathbf{x}))$. 

We handle the nonlinear transition function $f(\cdot)$ with the standard EKF prediction \cite{ribeiro2004kalman}, i.e.,
\begin{equation}
\begin{alignedat}{2}
\hat{\mathbf x}_{k\mid k-1}
&=
f\!\left(\hat{\mathbf x}_{k-1\mid k-1}\right),
\\
\mathbf P_{k\mid k-1}
&=
\mathbf{F}_k
\mathbf P_{k-1\mid k-1}
&&
\mathbf{F}_k^\top
{}+\mathbf Q\!\left(
\hat{\mathbf x}_{k-1\mid k-1}
\right),
\end{alignedat}
\end{equation}
where \(\hat{\mathbf x}_{k-1\mid k-1}\) and \(\hat{\mathbf x}_{k\mid k-1}\) denote the corrected estimate and one-step prediction, respectively, \(\mathbf P_{k-1\mid k-1}\) and \(\mathbf P_{k\mid k-1}\) are their covariances, and \( \mathbf{F}_k=\nabla f(\hat{\mathbf x}_{k-1\mid k-1})\) is the transition Jacobian of \(f\) evaluated at \(\hat{\mathbf x}_{k-1\mid k-1}\). The transition function \(f(\mathbf x)\) has an important property. Specifically, for every \(\mathbf x\in[0,1]^{|\bar{\mathcal E}|+|\bar{\mathcal T}|}\) and every \(e\in\partial t\),
$
0\leq [f^2(\mathbf x)]_t
\leq [f^1(\mathbf x)]_e\leq 1,
$ since every entry of the probability vectors \(\mathbf p^1(\mathbf x)\) and \(\mathbf p^2(\mathbf x)\) is a convex combination of \(\beta^i\) and \(\rho^i\) and thus lies in \([0,1]\). Hence, \(f(\mathbf x)\in\mathcal C\), and since every corrected estimate is confined to \(\mathcal C\) (Sec.~\ref{sec:constrainted_update}), every prediction lies in $\mathcal C$.

\subsection{Measurement Equation}
\label{ssec:measurement}
We construct virtual measurements directly from the observed signals,
without requiring explicit graph-filter input-output pairs as in
\cite{dabush2025sparsity,tenorio2025tracking,azoulay2026dynamic}. To relate the observed signals and the topology, we follow common assumptions in graphs and SC learning \cite{Kalofolias2016HowSignals,isufi2025topological}. We relate nodal observations to the edge topology by assuming smoothness across active edges. For an edge $e=\{a_e,b_e\}$, this smoothness measure is defined as
$
[\mathbf h_k^1]_e
=
\left\|
[\mathbf Z_k^0]_{a_e,:}
-
[\mathbf Z_k^0]_{b_e,:}
\right\|_2^2.
$
To relate edge signals to triangles, different measures can be used, such as low curl or similarity. For instance, we may assume that edge signals have low curl over active triangles \cite{isufi2025topological}. For a triangle $t$ with consistently oriented boundary edges indexed
by $i_t,j_t,\ell_t$, the low-curl cost can then be defined as
$
[\mathbf h_k^2]_t
=
\left\|
[\mathbf Z_k^1]_{i_t,:}
+
[\mathbf Z_k^1]_{j_t,:}
-
[\mathbf Z_k^1]_{\ell_t,:}
\right\|_2^2.$ Alternatively, we may use the orientation-invariant similarity cost
introduced in \cite{sarathchandran2026joint}, in which case $[\mathbf h_k^2]_t
=
\sum_{\substack{m,n\in\{i_t,j_t,\ell_t\}\\m<m}}
\left\|
[\mathbf Z_k^1]_{m,:}
-
[\mathbf Z_k^1]_{n,:}
\right\|_2^2.$ Thus, $\mathbf h_k^1$ and $\mathbf h_k^2$ collect the edge- and
triangle-level smoothness costs. Joint smoothness of the SC is encoded through the virtual measurement
\[
0
=
(\mathbf h_k^1)^\top\mathbf x_k^1
+
(\mathbf h_k^2)^\top\mathbf x_k^2
+
\nu_k^{\mathrm s},
\]
where the first two terms on the R.H.S. denote the topology-aware smoothness measures for the node and edge signals, respectively.

\noindent\textbf{Remark. } The virtual-measurement framework may be extended to other observation-only topology-signal
models, such as higher-order structural equation or
graph-Volterra models, by expressing their
topology-signal relations as measurement equations
\cite{Self-DrivenPrediction}.

Following sparsity-aware filtering methods that inject sparsity
information through additional measurements
\cite{farahmand2014tracking,carmi2009methods,hage2020sparse}, we also prescribe the
edge and triangle cardinalities $C_k^1$ and $C_k^2$. Stacking the
smoothness and cardinality measurements gives
\begin{equation}
\underbrace{
\begin{bmatrix}
0\\ C_k^1\\ C_k^2
\end{bmatrix}}_{\mathbf y_k}
=
\underbrace{
\begin{bmatrix}
(\mathbf h_k^1)^\top &(\mathbf h_k^2)^\top\\
\mathbf 1_{|\bar{\mathcal{E}}|}^\top&\mathbf 0_{|\bar{\mathcal{T}}|}^\top\\
\mathbf 0_{|\bar{\mathcal{E}}|}^\top&\mathbf 1_{|\bar{\mathcal{T}}|}^\top
\end{bmatrix}}_{\mathbf H_k}
\mathbf x_k
+
\boldsymbol\nu_k .
\label{eq:measurements}
\end{equation}
The measurement errors are independent, with
$\boldsymbol\nu_k\sim\mathcal N(\mathbf0,\mathbf R)$ and
$\mathbf R=\operatorname{diag}
(\sigma_{\mathrm s}^2,\sigma_{\mathrm c,1}^2,
\sigma_{\mathrm c,2}^2)$.
Here, $\sigma_{\mathrm s}^2$ captures uncertainty and mismatch in the
signal-topology model, whereas $\sigma_{\mathrm c,i}^2$ controls
adherence to the approximate count $C_k^i$. For the zero-valued smoothness measurement, the cardinality measurements prevent the trivial all-zero solution. Under a model such as a structural equation model, they instead introduce sparsity.
\subsection{Constraint-Aware Measurement Update}
\label{sec:constrainted_update}
Given the Gaussian predictive approximation and linear measurement
model, the maximum a posteriori (MAP) update of $\mathbf{x}_k$ is the Kalman update. Our updates, however, have to adhere to the feasible set $\mathcal C$. We may directly impose $\mathcal C$ via a constrained MAP estimate  \cite{simon2010kalman}.  Define the quadratic MAP objective and
its Hessian as
\[
\begin{aligned}
\mathcal L_k(\mathbf x)
={}&
\frac12
\left\|
\mathbf x-\hat{\mathbf x}_{k\mid k-1}
\right\|_{\mathbf P_{k\mid k-1}^{-1}}^2
+
\frac12
\left\|
\mathbf y_k-\mathbf H_k\mathbf x
\right\|_{\mathbf R^{-1}}^2,\\
\boldsymbol\Lambda_k
:={}&
\nabla^2\mathcal L_k(\mathbf x)
=
\mathbf P_{k\mid k-1}^{-1}
+
\mathbf H_k^\top\mathbf R^{-1}\mathbf H_k,
\end{aligned}
\]
where $\|\mathbf z\|_{\mathbf M}^2=\mathbf z^\top\mathbf M\mathbf z$.
The constrained MAP (CM) estimate is $\hat{\mathbf x}_{k\mid k}^{\mathrm{CM}}
=
\arg\min_{\mathbf x\in\mathcal C}
\mathcal L_k(\mathbf x).$ A common approximation is to constrain only the state estimate while the unconstrained covariance update is retained
\cite{simon2010kalman},
$\mathbf P_{k\mid k}^{\mathrm{CM}}
=
\boldsymbol\Lambda_k^{-1}.$

This covariance, however, ignores the geometry of $\mathcal C$. To let the constraints inform both the state and the covariance updates, we encode them as nonlinear zero-valued pseudo-measurements, building on similar inequality-constrained approaches \cite{husmann2025recursive}, as we detail next.

Let $\mathbf A_{\mathrm{inc}}\in
\mathbb R^{3|\bar{\mathcal T}|\times(|\bar{\mathcal E}|+|\bar{\mathcal T}|)}$ encode inclusion, with the row
associated with $(t,e)$, $e\in\partial t$, satisfying
$[\mathbf A_{\mathrm{inc}}\mathbf x]_{(t,e)}
=[\mathbf x^1]_e-[\mathbf x^2]_t$. Besides inclusion, we impose the
box constraint $\mathbf x_k\in[0,1]^{|\bar{\mathcal E}|+|\bar{\mathcal T}|}$. We impose these constraints via the following logarithmic barrier functions:
$B_{\mathrm{inc}}(\mathbf x)
=-\mathbf 1^\top
  \log(\mathbf A_{\mathrm{inc}}\mathbf x),
B_{\mathrm{box}}(\mathbf x)
=-\mathbf 1^\top
  \bigl[\log\mathbf x+\log(\mathbf 1-\mathbf x)\bigr],
$
 where all logarithms are element-wise. For
$\text{a}\in\mathcal A:=\{\mathrm{inc},\mathrm{box}\}$, we introduce the
virtual measurement
\begin{equation}   
\begin{aligned}
0=h_\text{a}(\mathbf x_k)+\nu_k^\text{a}, h_\text{a}(\mathbf x)&=\sqrt{B_\text{a}(\mathbf x)}, 
\nu_k^\text{a}\sim\mathcal N(0,\sigma_\text{a}^2).
\end{aligned}
\end{equation}
Appending these nonlinear measurements to the linear model in \eqref{eq:measurements} adds logarithmic barrier regularizers to the objective $\mathcal{L}_k(\mathbf{x})$:
\begin{equation}    
\begin{aligned}
\mathcal J_k(\mathbf x)
=
\mathcal L_k(\mathbf x)
+
\sum_{\text{a}\in\mathcal A}
\lambda_\text{a} B_\text{a}(\mathbf x),
\end{aligned}
\end{equation}
where $\operatorname{dom}\mathcal J_k=\operatorname{int}\mathcal C$, and $\lambda_\text{a}=(2\sigma_\text{a}^2)^{-1}.$ We next consider three progressively richer treatments of the nonlinear pseudo-measurements, differing in how their nonlinearities enter the state and covariance updates. The standard EKF uses a single linearization, the IEKF retains the nonlinear MAP correction with Gauss-Newton covariance information, and the Laplace approximation incorporates the full local curvature of the objective.
\begin{figure*}[t]
  \centering
  \includegraphics[width=\textwidth]{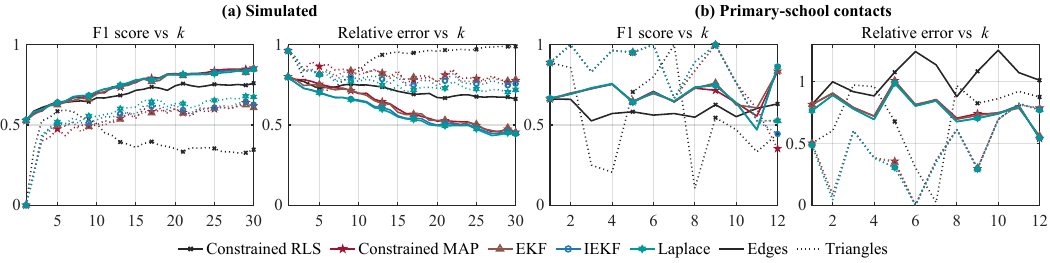}
  \caption{Thresholded support recovery and unthresholded topology error for
  (a) simulated data and (b) real data.}
  \label{fig:sim-vs-real}
\end{figure*}

\noindent\textbf{Standard EKF.}
The standard EKF linearizes each nonlinear measurement once at the
predicted estimate, yielding
\begin{equation}
\begin{aligned}
\tilde{\mathbf x}_{k\mid k}^{\, \text{EKF}}
=
\arg\min_{\mathbf x}
\Bigg\{
\begin{alignedat}{1}
\mathcal L_k(\mathbf x)
+\sum_{\text{a}\in\mathcal A}\lambda_\text{a}
\Big[
h_\text{a}(\hat{\mathbf x}_{k\mid k-1})
&\\[-1mm]
{}+
\nabla h_\text{a}(\hat{\mathbf x}_{k\mid k-1})^\top
(\mathbf x-\hat{\mathbf x}_{k\mid k-1})
\Big]^2
&
\end{alignedat}
\Bigg\},
\\
\left(\mathbf P_{k\mid k}^{\mathrm{EKF}}\right)^{-1}
=
\boldsymbol\Lambda_k
+
\sum_{\text{a}\in\mathcal A}2\lambda_\text{a}
\nabla h_\text{a}(\hat{\mathbf x}_{k\mid k-1})
\nabla h_\text{a}(\hat{\mathbf x}_{k\mid k-1})^\top .
\end{aligned}
\end{equation}
Because the affine update need not lie in $\mathcal C$, we project
$\tilde{\mathbf x}_{k\mid k}^{\mathrm{EKF}}$ using \(\hat{\mathbf x}^{\mathrm{EKF}}_{k\mid k}
=
\arg\min_{\mathbf x\in\mathcal C}
\left\|
\mathbf x-\tilde{\mathbf x}^{\mathrm{EKF}}_{k\mid k}
\right\|^2_{(\mathbf P^{\mathrm{EKF}}_{k\mid k})^{-1}}.\)

\noindent\textbf{Iterated EKF.}
An IEKF repeatedly linearizes the nonlinear measurement in $\mathcal{J}_k(\mathbf{x})$ at the
previous iterate and solves the resulting quadratic problem. If these iterations converge, their limit is a stationary point and hence, since $\mathcal{J}_k(\mathbf{x})$ is convex, a minimizer of $\mathcal{J}_k(\mathbf{x})$. Following \cite{farahmand2014tracking}, we therefore compute this minimizer directly as $\hat{\mathbf x}_{k\mid k}^{\text {IEKF}}
= \arg\min_{\mathbf x}\mathcal J_k(\mathbf x)$. Because the barriers remain in the objective, no subsequent projection
is required. 

The corresponding IEKF covariance uses the Gauss-Newton information
evaluated at the corrected rather than the predicted estimate \cite{farahmand2014tracking}:
\begin{equation}  
\begin{aligned}
\left(\mathbf P_{k\mid k}^{\mathrm{IEKF}}\right)^{-1}
=
\boldsymbol\Lambda_k
+\sum_{\text{a}\in\mathcal A}2\lambda_\text{a}
\nabla h_\text{a}(\hat{\mathbf x}_{k\mid k}^{\text {IEKF}})
\nabla h_\text{a}(\hat{\mathbf x}_{k\mid k}^{\text {IEKF}})^\top .
\end{aligned}
\end{equation}
\noindent\textbf{Laplace Approximation.}
The Laplace approximation represents a distribution locally by a Gaussian centered at its mode, with the curvature at that mode as its precision \cite{tierney1986accurate,fatemi2012study}. Thus the state update is the minimizer of $\mathcal{J}_k(\mathbf{x})$ and is the same as that in the IEKF, $\hat{\mathbf x}^{\mathrm{Lap}}_{k\mid k}
:=\hat{\mathbf x}^{\mathrm{IEKF}}_{k \mid k}$. The covariance update however, retains the complete Hessian  of $\mathcal{J}_k(\mathbf{x})$:
\begin{equation}
\begin{aligned}
\left(\mathbf P_{k\mid k}^{\mathrm{Lap}}\right)^{-1}
={}
\boldsymbol\Lambda_k
&+
\sum_{\text{a}\in\mathcal A}2\lambda_\text{a}
\Big[
\nabla h_\text{a}(\hat{\mathbf x}^{\mathrm{Lap}}_{k \mid k})
\nabla h_\text{a}(\hat{\mathbf x}^{\mathrm{Lap}}_{k|k})^\top\\
&\hspace{17mm}
+
h_\text{a}(\hat{\mathbf x}^{\mathrm{Lap}}_{k|k})
\nabla^2 h_\text{a}(\hat{\mathbf x}^{\mathrm{Lap}}_{k|k})
\Big].
\end{aligned}
\end{equation}

The first term in the parenthesis is the IEKF Gauss-Newton information, whereas the second restores the omitted second-order measurement curvature. Thus, CM constrains only the state estimate, whereas the EKF, IEKF,
and Laplace methods incorporate progressively richer constraint
information into the covariance via nonlinear measurements.
\section{Numerical Experiments}
We evaluate the proposed methods on simulated and real-world data, complete details of which are available in our \href{https://github.com/VarunSarathchandran/DynamicSimplicialComplexTracking.git}{GitHub repository}.
\subsection{Simulated Experiments}
\noindent\textbf{Setup.}
We initialize a seed SC with $N=15$ nodes from an Erdős--Rényi graph
and uniformly activate half its feasible triangles. We then roll out
\eqref{eq:intrinsic_dynamics}--\eqref{eq:triangle_gate} with
$\beta^1=\beta^2=0.05$ and $\rho^1=\rho^2=0.98$.
At each time $k$, smooth nodal and low-curl edge signals are generated by
Laplacian-filtering random inputs \cite{sarathchandran2026joint}.


As baselines, we use constrained RLS, obtained by setting
$f(\mathbf x)=\mathbf x$ and $\mathbf Q(\mathbf x)=\mathbf0$, and CM,
which, in the spirit of \cite{dabush2025sparsity}, enforces constraints only in the state correction. Dynamics-informed methods
receive the true transition probabilities, and all methods receive the
true simplex counts. From the relaxed estimates, we report support F1
under a common threshold and relative error before thresholding.
Note that thresholding preserves the inclusion constraints in $\mathcal{C}$.

\noindent{\textbf{Results.} Fig.~\ref{fig:sim-vs-real}(a) shows small differences in edge-support
F1 but clearer gains in triangle recovery and relative error. Laplace
generally performs best, followed by IEKF. Although their advantage over CM narrows in some instances, neither performs appreciably worse than it. EKF occasionally falls below CM, suggesting that
single linearization followed by projection can offset the benefit of
constraint-informed covariance. RLS performs worst because it ignores
the topology-transition model.
\subsection{Real Experiments}
\noindent\textbf{Setup.} We use interaction data from a primary
school \cite{Benson2018Simplicial,Stehle2011High}, which records the groups of students who interact at a 20-second resolution. We form time-varying SCs over 20-minute windows: an edge $(a,b)$ is active if $a$ and $b$ interact pairwise or co-occur in any higher-order interaction, and a triangle $(a,b,c)$ only if the three co-occur in a single second- or higher-order interaction. Inclusion therefore holds by construction, while triangles remain unfilled unless the group interaction actually took place: $a$, $b$ and $c$ may be pairwise
connected within a window without ever having met as a group. We track one classroom of 25 students over a 4 hour window. The dataset carries no signals, so we simulate smooth nodal signals and low-curl edge signals. The experiment is thus semi-real: a real topology, observed through synthetic signals that obey our model. Baselines in the simulated case are retained.

\noindent\textbf{Calibration.} Quantities assumed known in simulation are estimated here. The data spans two days and every parameter is learnt on day~1, frozen, and deployed on day~2. Birth and survival probabilities are the fractions of inactive (resp.\ active) simplices active at the next step, with
triangle transitions counted only over triples whose boundary edges are
present, yielding $\beta_1=0.16$, $\rho_1=0.50$, $\beta_2=0.13$,
$\rho_2=0.32$. The counts $C^1_k,C^2_k$ are those observed in the matching day-1 window, and the measurement variances are fixed on day~1 as regularisation weights rather than calibrated noise levels.

\noindent\textbf{Results.} Figure \ref{fig:sim-vs-real}(b) shows that all estimators outperform RLS on both metrics at every time index but two. Both exceptions are steps at which the calibrated simplex counts are confidently wrong. RLS, which leans more heavily on the prior, is less damaged by such a measurement. Differences among our own estimators are visible on relative error but marginal, and we leave to future work why the improved curvature information does not separate them as clearly as in simulation. The consistent margin over RLS, together with F1 scores seen in \ref{fig:sim-vs-real}(b), nonetheless show that the framework tracks a real time-varying topology when the observation model holds.
\section{Conclusion}
We introduced a state-space framework for jointly tracking the edges and triangles of a time-varying simplicial complex. We also introduced a closure-aware birth-survival model which provides the transition mean and covariance, while virtual measurements relate the latent topology to observed signals and encode its structural constraints. We investigated SC tracking based on an EKF, IEKF, and Laplace approximation, with numerical results demonstrating the value of topology-aware dynamics and constraint-informed covariance updates, particularly for simulated results. Future work will study observability and identifiability under limited measurements and develop scalable algorithms for larger candidate complexes.
\clearpage 
\noindent\textbf{Acknowledgment.}
This publication is part of the CYCLONE project
(file No.~OCENW.M.24.255) under the NWO Open Competition Domain
Science--M programme, partly financed by the Dutch Research Council
(NWO), DOI: \url{https://doi.org/10.61686/UBZOA14639}.
This work was also supported in part by the TU Delft AI Labs programme,
NWO OTP GraSPA proposal No.~19497, NWO Veni grant No.~222.032, and the
SURE-AI Centre under grant No.~357482 from the Research Council of
Norway. Claude (Anthropic) was used to assist with writing the MATLAB code. The authors take full responsibility for the results of this paper. \\

\noindent\textbf{Compliance with Ethical Standards.} This is a numerical simulation study for which no ethical approval was required.

\bibliographystyle{IEEEbib}
\bibliography{refs,references-2}
\end{document}